\documentclass[runningheads]{llncs}
\usepackage[T1]{fontenc}
\renewcommand{\arraystretch}{1.1}
\usepackage{url}
\usepackage{tikz}
\usepackage{booktabs}
\usepackage{cite}
\usepackage{amsmath,amssymb,amsfonts}
\usepackage{algorithmic}
\usepackage{textcomp}
\usepackage{xcolor}
\usepackage{pifont}
\newcommand{\cmark}{\ding{51}}
\newcommand{\xmark}{\ding{55}}

\def\BibTeX{{\rm B\kern-.05em{\sc i\kern-.025em b}\kern-.08em
    T\kern-.1667em\lower.7ex\hbox{E}\kern-.125emX}}
\usepackage{graphicx}
\begin{document}
\title{Binary and Multiclass Cyberattack Classification on GeNIS Dataset}%

\author{Miguel Silva\thanks{Corresponding author: Miguel Silva email: mdgsa@isep.ipp.pt}\inst{1}\orcidID{0009-0008-6630-9939} \and
Daniela Pinto\inst{1}\orcidID{0009-0000-3003-6694} \and
João Vitorino\inst{1}\orcidID{0000-0002-4968-3653} \and
Eva Maia\inst{1}\orcidID{0000-0002-8075-531X}\and
Isabel Praça\inst{1}\orcidID{0000-0002-2519-9859}and
Ivone Amorim\inst{1}\orcidID{0000-0001-6102-6165}\and
Maria João Viamonte\inst{1}\orcidID{0000-0002-6464-3529}}
\authorrunning{M. Silva et al.}
%
\institute{GECAD, ISEP, Polytechnic of Porto, rua Dr. António Bernardino de Almeida, 4249-015 Porto, Portugal}
\maketitle

\begin{abstract}
The integration of Artificial Intelligence (AI) in Network Intrusion Detection Systems (NIDS) is a promising approach to tackle the increasing sophistication of cyberattacks. However, since Machine Learning (ML) and Deep Learning (DL) models rely heavily on the quality of their training data, the lack of diverse and up-to-date datasets hinders their generalization capability to detect malicious activity in previously unseen network traffic. This study presents an experimental validation of the reliability of the GeNIS dataset for AI-based NIDS, to serve as a baseline for future benchmarks. Five feature selection methods, Information Gain, Chi-Squared Test, Recursive Feature Elimination, Mean Absolute Deviation, and Dispersion Ratio, were combined to identify the most relevant features of GeNIS and reduce its dimensionality, enabling a more computationally efficient detection. Three decision tree ensembles and two deep neural networks were trained for both binary and multiclass classification tasks. All models reached high accuracy and F1-scores, and the ML ensembles achieved slightly better generalization while remaining more efficient than DL models. Overall, the obtained results indicate that the GeNIS dataset supports intelligent intrusion detection and cyberattack classification with time-based and quantity-based behavioral features.
\end{abstract}

\section{Introduction}
The growing frequency and sophistication of cyberattacks make traditional Network Intrusion Detection Systems (NIDS), using signature-based detection, to be increasingly ineffective at identifying emerging or previously unseen threats \cite{traditional_ids}. To address this problem, NIDS are being incorporated with Artificial Intelligence (AI) due to its ability to learn from historical data and classify unknown data based on previous patterns. However, training robust and reliable Machine Learning (ML) and Deep Learning (DL) models requires access to accurate and representative data reflecting current network services, protocols, and behaviors. Since existing datasets lack data diversity and contain network traffic that is becoming outdated, it is increasingly difficult to detect modern threats \cite{outdated, outdated2}. In addition to becoming outdated, recent studies have also shown that many publicly available datasets contain inconsistencies caused by flaws in their feature extraction processes \cite{flaws}. Despite significant efforts to correct these datasets and release improved versions, discrepancies still arise. This is mainly due to variations in flow exporters used to process the same raw network captures that initially introduced these errors \cite{Hera2025}. 

This study presents the first experimental validation and evaluation of GeNIS \cite{genis}, a recently developed dataset designed for network traffic analysis and AI-based NIDS in small to medium sized organizations. After a careful analysis of the modular scenarios of GeNIS, five feature selection techniques were employed to better understand the dataset and reduce its dimensionality: Information Gain, Chi-Squared Test, Recursive Feature Elimination, Mean Absolute Deviation, and Dispersion Ratio. Lastly, the reliability of GeNIS for binary and multiclass classification was evaluated with several ML and DL models, considering standard evaluation metrics and their training and inference times.

The ML models considered for this study were decision tree ensemble models commonly used in NIDS: Random Forest (RF), Extreme Gradient Boosting (XGB), and Light Gradient Boosting Machine (LGBM). In turn, the more complex DL models followed two distinct architectures: Long Short-Term Memory (LSTM) and Multilayer Perceptron (MLP). Different versions of each model were trained on the full feature set and a subset of selected features derived from selection techniques. To ensure that each model behaved as expected, the SHapley Additive exPlanations (SHAP) \cite{shap} technique was applied and the most impactful features of GeNIS were analyzed.

This paper is structured into multiple sections. Section~\ref{related_work} reviews issues identified in previous datasets and summarizes recent ML and DL results. Section~\ref{dataset} introduces the GeNIS dataset and includes a brief exploratory data analysis. Section~\ref{experimental} describes the experimental setup, presents and discusses the obtained results of binary and multiclass classification, followed by a discussion of the models' performance and key findings. Finally, Section~\ref{conclusion} presents the main conclusions and future work.

\section{Related Work}
\label{related_work}

To protect organizational networks from malicious actors and develop robust NIDS that can secure critical assets, numerous studies have produced datasets that are commonly used to train AI models, particularly ML and DL approaches, to identify malicious patterns in network traffic. Creating such datasets requires several key steps, including collecting, extracting, and analyzing traffic data, which can be done using integrated solutions or a combination of specialized tools. However, inconsistencies and incompatibilities among these tools can lead to discrepancies. Different tools may extract different values for the same feature from the same raw data, which can introduce errors into the final dataset \cite{PINTO2025111177}.

A prominent example is the CICIDS2017 dataset \cite{cicids}, which is widely used in benchmarking studies \cite{benchmark_cicids2017,benchmark_cicids2017_2}. Due to its popularity, several researchers have examined this dataset and identified significant inaccuracies, prompting the release of corrected versions. Specifically, Engelen et al. \cite{engelen} and Liu et al. \cite{engelen_2} discovered multiple issues in the dataset’s creation process. They reported a misimplementation of the DoS Hulk attack and identified flaws in CICFlowMeter\footnote{\url{https://www.unb.ca/cic/research/applications.html}}, the tool used for flow generation. These flaws included errors in feature extraction and a fundamental misunderstanding of the Transmission Control Protocol (TCP). The dataset also suffered from label inaccuracies and corrupted attack samples.

Further analysis by Rosay et al. \cite{rosay} revealed issues in the CSV files provided by the dataset authors. These issues were not present in the raw traffic captures, suggesting that they originated from CICFlowMeter. Consequently, concerns have been raised that other datasets generated with the same tool may also contain similar flaws. Later, Lanvin et al. \cite{Lanvin} identified even more issues than previously reported, including incorrectly labeled port scan attacks and duplicated traffic, which make feature extraction and labeling more difficult.

A more recent dataset, HIKARI2021 \cite{hikari} , was developed using the same set of features as CICIDS2017, but it employed Zeek\footnote{\url{https://zeek.org/}} as the flow exporter. As with CICIDS2017, researchers have identified problematic features in HIKARI2021 that negatively impact the performance of ML models trained on the dataset. These problematic features are believed to act as record identifiers or reflect indexing errors introduced by the processing software \cite{fernandes}. The dataset was originally released with four category labels: Benign, BruteForce, Probing, and CryptoMiner. However, the authors subsequently released a newer version of the dataset \cite{new_hikari}, which included two additional PCAP files containing attacks already present in the earlier version, along with a revised labeling scheme. Notably, the Probing class disappeared in the updated version \cite{hikari_22_analysis}.

Recent efforts have focused on standardizing the dataset creation process for NIDS by introducing the all-in-one HERA flow exporter. HERA was developed in response to issues identified with CICFlowMeter, which has been linked to errors in several popular datasets, being able to generate flow records and extract relevant features consistently and reliably \cite{Hera2025}.

Following the creation of HERA and since prior datasets offer so many challenges, the GeNIS dataset was created specifically for AI-based NIDS applications. GeNIS is a modular dataset designed specifically for AI-based NIDS applications. With its focus on reflecting the typical user behavior, services, and network protocols of small to medium sized organizations, GeNIS aims to enhance the detection capabilities of these systems. This dataset offers a diverse range of benign and malicious traffic types, and its modular design allows organizations to tailor the dataset to their unique network scenarios, either by selecting specific subsets for targeted protection or by using the full dataset for broader, general-purpose defense.

To evaluate the real-world effectiveness and practicality of GeNIS in organizational networks, it is essential to validate the performance of AI models trained on it. Previous studies have used datasets such as CICIDS2017 to explore various deep learning models, including Deep Neural Networks, LSTM networks, and Convolutional Neural Networks, while reporting a binary classification performance accuracy of over 94\% using these models \cite{lstm_rw}. Similarly, LSTM models have demonstrated strong results on older datasets, such as KDD99, achieving over 95\% accuracy and F1-scores for binary classification across different model configurations, both with and without feature selection. For multiclass classification, the performance remained high with F1-scores above 91\% \cite{lstm_rw2}.

Studies using MLP architectures on datasets such as UNSW-NB15 have also achieved strong performance in multiclass settings, reporting weighted average F1-scores and accuracies around 82\%. These models were further improved by applying feature selection techniques \cite{mlp}. Similarly, tree-based ensemble models, such as LGBM, XGB, and RF, have demonstrated effective performance in binary classification tasks. In recent studies, these models consistently achieve F1-scores and accuracy above 80\%, both with the full feature set and after dimensionality reduction through feature selection \cite{arvores,arvores2}.

Overall, recent studies on network intrusion detection have achieved promising results with both traditional, tree-based ensemble models and more complex, DL neural network architectures. These models tend to maintain or enhance their performance when trained on reduced feature subsets, which highlights the effectiveness of feature selection techniques. However, due to significant flaws discovered in widely used datasets, it is crucial to evaluate these models' and feature selection methods' performance on more recent, reliable datasets free from such anomalies.


\section{Dataset Analysis}
\label{dataset}

The GeNIS dataset consists of eight attack scenarios and three benign activity scenarios. Malicious activity was performed using two Kali machines located in different parts of the network. Each attack scenario consists of multiple stages in which the attacker exploits network configurations, identifies and maps network assets, and disrupts normal business operations by launching Denial of Service (DoS) attacks or attempting to compromise administrator credentials with commonly used passwords. These attacks target services such as File Transfer Protocol (FPT), web servers, and Active Directory servers, while also generating network congestion.

The benign scenarios are divided into three distinct categories. The user activity scenario simulates typical employee behavior during working hours (9:00 a.m. to 4:00 p.m.), including common tasks such as web browsing, emailing, and system updates. The admin activity scenario represents typical system administrator behavior, including actions such as Secure Shell Protocol (SSH) access and system update maintenance, which also occur during standard working hours. Finally, the background activity scenario captures periods of no user interaction and represents idle traffic collected over the weekend.

A total of 125 features were extracted from the raw recordings of the GeNIS dataset using the HERA flow exporter. Three of these features are label-related and suitable for binary and multiclass classification tasks. The \textit{BinaryLabel} distinguishes between \textit{benign} and \textit{malicious} traffic; the \textit{CategoryLabel} provides a general classification, separating traffic into categories such as \textit{benign}, \textit{bruteforce}, \textit{DoS}, and \textit{reconnaissance}, also known as \textit{recon}. The \textit{SubCategoryLabel} offers a more granular view by differentiating between three benign scenarios, identifying the specific protocols involved in \textit{bruteforce} and \textit{recon} attacks, and distinguishing the types of \textit{DoS} attacks carried out by attackers.

The remaining features can be grouped into five categories according to their nature. General features include basic identifiers, protocol information, and metadata. Time-based features include the timestamp marking the start of a flow, the duration of inactivity between packets, and inter-arrival times, all of which offer insights into the temporal aspects of network behavior. Quantity-based features include packet counts, packet sizes, and byte volumes, which are measured separately for each flow's source (the sender) and destination (the responder). Meanwhile, hybrid features combine time and quantity aspects, and context-based features reflect the specific characteristics of the network topology. 

Among the 17 general features, 6 (\textit{FlowID}, \textit{Rank}, \textit{Seq}, \textit{AutoId}, \textit{TcpOpt}, and \textit{Cause}) are generated by Argus and do not meaningfully contribute to understanding traffic behavior. Among the 38 quantity-based features, \textit{Ssaddr} and \textit{Sdaddr} are specific to HERA, representing the number of connections involving the same service and source/destination address. While these features are potentially useful for identifying attacks, they are highly dependent on the network’s topology and attacks, and should be excluded to ensure model generalization across different environments. Similarly, some of the 29 context-based features may encode specific network identifiers, including IP addresses, MAC addresses, and VLAN IDs. These values should be excluded to prevent the model from overfitting particular network setups rather than learning behavioral patterns.


Initial pre-processing of the GeNIS dataset was performed to ensure that no empty rows or columns were present, and the pre-processed dataset was provided alongside the original and raw traffic captures. Additionally, categorical features such as \textit{State}, \textit{Flags}, and \textit{Protocol} were one-hot encoded, resulting in 87 total features. While this version of the dataset is ready for direct use in training and testing models, it is important to note that the pre-processing was performed manually. Consequently, the resulting feature set may not be optimal for the final model's performance. To address this issue, statistical feature selection methods must be applied to identify the most relevant features and reduce dimensionality. Table \ref{tab:tab1} presents the number of flows extracted in 60-second intervals, grouped by \textit{BinaryLabel} and \textit{CategoryLabel}.

\begin{table}[h]
\caption{Flows per class}
\begin{center}
\begin{tabular}{@{}lllll@{}}
\toprule
\textbf{Type} & \textbf{Class Label} & \textbf{Train} & \textbf{Test} & \textbf{Ratio (\%)} \\ \midrule
Binary & Malicious & 273124 & 68282 & 92.63\% \\
 & Benign & 21720 & 5430 & 7.37\% \\ \midrule
Multiclass & DoS & 236512 & 59128 & 80.22\% \\
 & Recon & 22186 & 5547 & 7.52\% \\
 & Benign & 21720 & 5430 & 7.37\% \\
 & Bruteforce & 14426 & 3607 & 4.89\% \\ \bottomrule
\end{tabular}
\label{tab:tab1}
\end{center}
\end{table}

\section{Experimental Findings}
\label{experimental}

This section outlines the feature selection methods and models evaluated, along with the results obtained. All experiments were conducted on a standard laptop with 16GB of RAM, a 6-core CPU and 8GB GPU. The implementation was developed in Python and the following libraries were used: \textit{numpy} and \textit{pandas} for general data manipulation, \textit{xgboost} for the implementation of XGB, \textit{lightgbm} for LGBM, \textit{tensorflow} for building the LSTM and MLP, and \textit{scikit-learn} for the implementation of RF, the StandardScaler, and feature selection.

\subsection{Experimental Setup}

In order to evaluate the models' ability to generalize effectively and make accurate predictions on the GeNIS dataset, features that could allow the models to infer classes based on specific network topology or attack methodology were excluded. This step was essential to ensure that the models learned behavioral patterns instead of memorizing structural or environment-specific indicators. Five feature selection methods were applied to identify and rank the most impactful features, targeting both binary and multiclass classification objectives. These methods include Information Gain, Chi-Squared Test, Recursive Feature Elimination, Mean Absolute Deviation, and Dispersion Ratio. The results from each feature selection method were first normalized and then aggregated by summing the corresponding scores for each feature. Lastly, the 16 most significant features with the highest combined scores were selected. A description of each technique is provided below.

\textit{Information Gain}. An entropy-based measure that quantifies the reduction in uncertainty about the target class after observing a given feature \cite{information_gain}.

\textit{Chi-Squared Test}. A statistical method that evaluates the strength of the association between each feature and the target class, by selecting features that exhibit a high level of dependency on class labels \cite{chi_squared}.

\textit{Recursive Feature Elimination}. An iterative technique that fits a model and recursively removes the least important feature based on the model's internal importance scores to gradually reduce the set of relevant features \cite{rfe}. 

\textit{Mean Absolute Deviation}. A statistical measure of variability obtained by calculating the average of the absolute differences between each value and the mean. Features with higher divergence are considered more informative \cite{mad}.

\textit{Dispersion Ratio}. A measure based on the calculation of the square root of the ratio between the sum of the squared deviations and the overall dispersion within the dataset \cite{dr}.

Three ML models based on decision tree ensembles, RF, XGB, and LGBM, and two DL models based on deep neural networks, LSTM, and MLP, were considered. The optimal configuration for each model was determined through a grid search involving an exhaustive evaluation of combinations of hyperparameters using 5-fold cross-validation. This process was applied to the full set of features as well as to reduced sets obtained through feature selection for both binary and multiclass classification tasks.

Performance was assessed using the F1-score for binary classification and the macro-averaged F1-score for multiclass classification to ensure balanced evaluation across all classes. The configurations that performed best during the tuning process were selected for final testing. The models and their respective fine-tuned hyperparameters are detailed below.

\textit{Random Forest}. RF \cite{rf_author} works by training a collection of decision trees on different parts of the data. Instead of relying on a single tree, it chooses the most frequent prediction from multiple trees. Table~\ref{tab:rf-params} summarizes the configuration.

\begin{table}[h]
\caption{Summary of RF configuration}
\label{tab:rf-params}
\begin{center}
\begin{tabular}{@{}ll@{}}
\toprule
\textbf{Parameter} & \textbf{Value} \\
\midrule
Criterion & Gini Impurity \\
Number of estimators & 100 \\
Maximum features & $\sqrt{\text{Number of features}}$\\
Maximum tree depth & 16 \\
Minimum samples in a leaf & 1\\

\bottomrule
\end{tabular}
\end{center}
\end{table}

\textit{Extreme Gradient Boosting}. XGB \cite{xgb_author} is a gradient boosting ensemble that combines the results of multiple decision trees, with each new tree correcting the errors of the previous ones. To identify optimal data splits, a histogram-based technique was used to reduce memory usage. The configuration is summarized in Table~\ref{tab:xgb-params}.

\begin{table}[h]
\caption{Summary of XGB configuration}
\label{tab:xgb-params}
\begin{center}
\begin{tabular}{@{}ll@{}}
\toprule
\textbf{Parameter} & \textbf{Value} \\
\midrule
Method & Histogram \\
Minimum loss reduction & 0.01 \\
Number of estimators & 100 \\
Learning rate & 0.2 \\
Maximum tree depth & 4 to 16 \\ 
Feature subsample & 0.8 to 0.9\\
\bottomrule
\end{tabular}
\end{center}
\end{table}

\textit{Light Gradient Boosting Machine}. LGBM \cite{lgbm_author} is another boosting ensemble that uses Gradient-based One-Side Sampling (GOSS) to build the decision trees, which is computationally lighter than the other approaches. The configuration of this model is summarized in Table~\ref{tab:lgbm-params}.

\begin{table}[h]
\caption{Summary of LGBM configuration}
\label{tab:lgbm-params}
\begin{center}
\begin{tabular}{@{}ll@{}}
\toprule
\textbf{Parameter} & \textbf{Value} \\
\midrule
Method & GOSS \\   
Number of estimators & 100 \\
Minimum loss reduction & 0.01 \\
Learning rate & 0.05 \\
Maximum leaves in a tree & 15 \\
Minimum samples in a leaf & 2 to 4 \\
Feature subsample & 0.8 \\
\bottomrule
\end{tabular}
\end{center}
\end{table}

As opposed to using automated hyperparameter optimization, the number of units in each hidden layer of both neural networks was manually adjusted based on experimental results. Two architectures were compared: one with 128 and 64 units and another with 64 and 32 units. To prevent overfitting, early stopping was applied during training to automatically restore the model weights from the epoch with the best performance, measured by loss and accuracy. All numerical input features were standardized using the StandardScaler pre-processing method, which scales features to have a mean of zero and a standard deviation of one. From the training data, 70\% of the dataset was used to train the models, and 30\% was used for validation.

\textit{Long Short-Term Memory}. LSTM \cite{lstm_author} is a specialized form of recurrent neural network that excels at modeling long-range dependencies in sequential data. Since this dataset is not inherently sequential, each data point is treated as one one-step sequence, enabling LSTMs to reveal complex nonlinear feature interactions. As outlined in Table~\ref{tab:lstm-params}, the model structure includes one input layer, two hidden layers (an LSTM layer followed by a Dense layer), and one output layer.

\begin{table}[h]
\caption{Summary of LSTM configuration}
\label{tab:lstm-params}
\begin{center}
\begin{tabular}{@{}ll@{}}
\toprule
\textbf{Parameter} & \textbf{Value} \\
\midrule
First layer units & 64 to 128\\
Second layer units & 32 to 64\\
Dropout & 0.2 \\
Optimizer & Adam \\
Learning rate & 0.001 \\
Batch size & 32 \\
Epochs & 30 \\
Early stopping patience & 3  \\ \bottomrule
\end{tabular}
\end{center}
\end{table}

\textit{Multilayer Perceptron}. MLP \cite{mlp_author} consists of multiple layers of neurons, including an input layer, one or more hidden layers, and an output layer. Each neuron in a layer is connected to every neuron in the next layer, allowing the network to learn complex nonlinear mappings between inputs and outputs through weighted connections and nonlinear activation functions. The specific model architecture is summarized in Table~\ref{tab:mlp-params}, featuring two Dense hidden layers.

\begin{table}[h]
\caption{Summary of MLP configuration}
\label{tab:mlp-params}
\begin{center}
\begin{tabular}{@{}ll@{}}
\toprule
\textbf{Parameter} & \textbf{Value} \\
\midrule
First layer units & 64 to 128\\
Second layer units & 32 to 64\\
Dropout & 0.2 \\
Optimizer & Adam \\
Learning rate & 0.001 \\
Batch size & 32 \\
Epochs & 30 \\
Early stopping patience & 3  \\ \bottomrule
\end{tabular}
\end{center}
\end{table}

To evaluate the performance of each model on the testing set, the following metrics were considered: F1-score (F1S), accuracy (ACC), recall (RCL), precision (PRC), false positive rate (FPR), average time per epoch (TE), training time (TT), and inference time (IT). Optimal performance would be represented as a score of 100\% on F1S, ACC, RCL, and PRC, a score of 0\% on FPR, and the lowest possible values for TT and IT metrics. The same metrics were applied to multiclass classification, with F1S, RCL, and PRC computed as macro averages to ensure equal consideration of each class.

Finally, an explainability technique was used to ensure that each model behaved as expected and that the selected features were contributing to a reliable detection. The SHAP technique was applied to analyze the predictions of each model and provide a score for each feature. These scores enabled the identification of which features were more relevant and the analysis of the combined impact of the different time, quantity, and hybrid-based behavioral features of the GeNIS dataset.

\subsection{Binary classification results}

In the binary classification task, the feature selection methods identified two features as the most significant, which together accounted for 56\% of the total feature importance. Based on these results, a final subset of 16 features was selected from the original set, representing 70\% of the cumulative importance. The selected features are listed in Table~\ref{tab:bin-fs}.

\begin{table}[h]
\caption{Feature selection for binary classification}
\label{tab:bin-fs}
\begin{center}
\renewcommand{\arraystretch}{1.2} 
\begin{tabular}{@{}l|llll@{}}
\hline
\textbf{Feature Type} & \multicolumn{4}{l}{\rule{0pt}{1.2em}\textbf{Selected Features}} \\ \hline
\textbf{Quantity-based} & DstTCPBase & SrcTCPBase & DstWin & SrcWin \\
 & TotBytes & TotPkts &  &  \\ \hline
\textbf{Time-based} & Dur & Min & Mean & RunTime \\
 & Sum &  &  &  \\ \hline
\textbf{Hybrid-based} & DstLoad & Load & Rate & SrcLoad \\
 & SrcRate &  &  &  \\ \hline
\end{tabular}
\end{center}
\end{table}

In the binary classification scenario, models trained using the full feature set demonstrated strong overall performance on all non-time-based metrics, with most models achieving near-perfect scores. The tree-based ensemble models slightly outperformed the neural networks in terms of F1S, while the neural networks matched or surpassed them in terms of FPR. Regarding TT, the ensemble models required less than half the TT of the neural networks and predicted the test set approximately ten times faster.

When using a reduced subset of features, there was a slight decline in performance across non-time-based metrics. However, TT for the ensemble models was reduced by approximately half. This reduction was not observed for the neural networks, as their training process differs, although they did benefit from shorter average TT per epoch. All models showed faster inference with the smaller feature set. However, a notable drawback was the significant increase in FPR. Some doubled their FPR, while others, like the MLP, experienced nearly a tenfold increase. The results obtained are described in Table~\ref{tab:bin-results}, where the best results between the model using all features and the smaller subset are highlighted.

\begin{table}[h]
\caption{Summary of binary classification results}
\label{tab:bin-results}
\begin{center}
\begin{tabular}{@{}llllllllll@{}}
\toprule
    \textbf{Model} & \textbf{FS} & \textbf{F1S} & \textbf{ACC} & \textbf{RCL} & \textbf{PRC} & \textbf{FPR} & \textbf{TT}  & \textbf{TE} & \textbf{IT} \\ \midrule
RF & \xmark & \textbf{99.9949} & \textbf{99.9905} & \textbf{100.0000} & \textbf{99.9897} & \textbf{0.1289} & 29.09 & - & 0.15 \\
 & \cmark & 99.9868 & 99.9756 & 99.9897 & 99.9839 & 0.2026 & \textbf{15.72} & - & \textbf{0.10} \\ \midrule
XGB & \xmark & \textbf{99.9949} & \textbf{99.9905} & \textbf{100.0000} & \textbf{99.9897} & \textbf{0.1289} & 4.78 & - & 0.06 \\
 & \cmark & 99.9714 & 99.9471 & 99.9927 & 99.9502 & 0.6262 & \textbf{1.81} & - & \textbf{0.05} \\ \midrule
LGBM & \xmark & \textbf{99.9905} & \textbf{99.9824} & \textbf{99.9927} & \textbf{99.9883} & \textbf{0.1473}  & 3.37 & - & 0.09 \\
 & \cmark & 99.9868 & 99.9756 & 99.9897 & 99.9839 & 0.2026 & \textbf{1.15} & - & \textbf{0.05} \\ \midrule
LSTM & \xmark & \textbf{99.9883} & \textbf{99.9783} & \textbf{99.9868} & \textbf{99.9897} & \textbf{0.1289} & \textbf{104.58} & 17.43 & 3.64 \\
 & \cmark & 99.8087 & 99.6459 & 99.6954 & 99.9222 & 0.9761 & 200.84 & \textbf{14.34} & \textbf{3.11} \\ \midrule
MLP & \xmark & \textbf{99.9883} & \textbf{99.9783} & \textbf{99.9854} & \textbf{99.9912} & \textbf{0.1105} & \textbf{92.82} & 11.60 & 2.75 \\
 & \cmark & 99.7954 & 99.6215 & 99.6719 & 99.9193 & 1.0129 & 125.69 & \textbf{11.43} & \textbf{2.53} \\ \bottomrule
\end{tabular}
\end{center}
\end{table}


SHAP analysis of binary classification models using a reduced feature subset reveals that predictions were driven primarily by quantity-related features. Unlike the other ensemble models, the RF model, similar to the LSTM and MLP neural networks, placed some emphasis on time-based features. However, RF uniquely gave greater attention to hybrid features than the other models did. Overall, the selected features included roughly an equal number of each feature type, and the quantity-based features provided a strong foundation for the models. This enabled the models to accurately classify most flows with only a slight decrease in performance. These results are summarized in Table~\ref{tab:bin-shap}.

\begin{table}[h]
\caption{Binary feature importance}
\label{tab:bin-shap}
\begin{center}
\begin{tabular}{@{}llll@{}}
\toprule
\textbf{Model} & \textbf{Quantity-based} & \textbf{Time-based} & \textbf{Hybrid-based} \\ \midrule
RF   & 1.22 & 0.26 & 0.52 \\
XGB  & 1.30 & 0.05 & 0.65 \\
LGBM & 1.75 & 0.02 & 0.23 \\
LSTM & 1.51 & 0.35 & 0.14 \\
MLP  & 1.28 & 0.54 & 0.18 \\ \bottomrule
\end{tabular}
\end{center}
\end{table}



\subsection{Multiclass classification results}

A similar pattern emerged regarding the features selected by the five feature selection methods applied to the multiclass dataset. The same two features ranked at the top, accounting for 56\% of the total feature importance together. Again, as in the binary classification, a subset of the top 16 features, representing 70\% of the cumulative importance, was selected. While the number of features and overall representativeness remained the same, four features were replaced with new ones in this multiclass subset. Table~\ref{tab:multi-fs} lists the features selected and used in this reduced subset.

\begin{table}[h]
\caption{Feature selection for multiclass classification}
\label{tab:multi-fs}
\begin{center}
\renewcommand{\arraystretch}{1.2} 
\begin{tabular}{@{}l|llll@{}}
\hline
\textbf{Feature Type} & \multicolumn{4}{l}{\rule{0pt}{1.2em}\textbf{Selected Features}} \\ \hline
\textbf{Quantity-based} & DstTCPBase & SrcTCPBase & DstWin & TotBytes \\
 & DstBytes & SAppBytes & SrcBytes & SrcWin \\ \hline
\textbf{Time-based} & Mean & Max & Sum & Dur \\
 & Min & RunTime &  &  \\ \hline
\textbf{Hybrid-based} & DstLoad & SrcLoad &  &  \\ \hline
\end{tabular}
\end{center}
\end{table}

Within the multiclass scenario, the performance of each model exhibited a pattern similar to that observed in the binary classification task. Overall evaluation metrics indicate that the models effectively predict whether flows represent attacks and distinguish between different attack types. Tree ensemble models required less TT and generally achieved higher F1S, although they did not always have the lowest false positive rate. Notably, the MLP predicted benign classes at rates comparable to XGB and LGBM, with the same FPR.

Using the smaller subset of features reduced both the TT and the IT of the three ML models on the test set. In contrast, the DL models required significantly more TT: nearly three times more for the MLP and more than one and a half times more for the LSTM. The MLP showed minimal improvement in inference time with the smaller feature set, while the LSTM demanded more time using less features. Switching from the full feature set to the reduced subset slightly increased the FPR for the tree-based models. Specifically, LGBM maintained the same number of correctly predicted benign flows, while the DL models experienced a substantial increase in FPR - between seven and nine times higher. However, these increases corresponded to less than 1\% (approximately 54 flows). Detailed results for each model are presented in Table~\ref{tab:multi-results}.

\begin{table}[h]
\caption{Summary of multiclass classification results}
\label{tab:multi-results}
\begin{center}
\begin{tabular}{@{}llllllllll@{}}
\toprule
\textbf{Model} & \textbf{FS} & \textbf{F1S} & \textbf{ACC} & \textbf{RCL} & \textbf{PRC}& \textbf{FPR} & \textbf{TT}  & \textbf{TE} & \textbf{IT} \\ \midrule
RF & \xmark & \textbf{99.9817} & \textbf{99.9919} & \textbf{99.9700} & \textbf{99.9933} & \textbf{0.0921} & 25.6 & - & 0.21 \\
 & \cmark & 99.9791 & 99.9905 & 99.9654 & 99.9929 & 0.1105 & \textbf{18.7} & - & \textbf{0.14} \\ \midrule
XGB & \xmark & \textbf{99.9817} & \textbf{99.9919} & \textbf{99.9724} & \textbf{99.9910} & \textbf{0.1105} & 9.51 & - & 0.10 \\
 & \cmark & 99.9175 & 99.9715 & 99.9052 & 99.9300 & 0.3499 & \textbf{4.95} & - & \textbf{0.07} \\ \midrule
LGBM & \xmark & \textbf{99.9755} & \textbf{99.9891} & \textbf{99.9650} & \textbf{99.9859} & 0.1105 & 5.15 & - & 0.27 \\
 & \cmark & 99.9687 & 99.9837 & 99.9633 & 99.9740 & 0.1105 & \textbf{3.00} & - & \textbf{0.20} \\ \midrule
LSTM & \xmark & \textbf{99.9659} & \textbf{99.9851} & \textbf{99.9555} & \textbf{99.9763} & \textbf{0.1289} & \textbf{210.05} & \textbf{17.50} & 3.58 \\
 & \cmark & 99.3162 & 99.7314 & 99.4644 & 99.1688 & 0.7182 & 335.28 & 17.66 & \textbf{3.48} \\ \midrule
MLP & \xmark & \textbf{99.9318} & \textbf{99.9729} & \textbf{99.9371} & \textbf{99.9266} & \textbf{0.1105} & \textbf{68.67} & \textbf{11.44} & \textbf{2.70} \\
 & \cmark & 98.7277 & 99.4791 & 99.3593 & 98.1148 & 0.9208 & 161.24 & 11.52 & 2.71 \\ \bottomrule
\end{tabular}
\end{center}
\end{table}

SHAP analysis revealed that the multiclass models using the smaller subset of features were primarily influenced by quantity-based features, as were the binary models. DL models relied more heavily on time-related features than ensemble models did, with LSTM placing relatively more importance on time features than other models, despite the fact that quantity features remained the primary drivers of its predictions. Nearly all models assigned similar levels of importance to hybrid features, despite notable differences in their proportions of time and quantity-based components, particularly between RF and XGB. Overall, these results suggest that models tasked with classifying attack types tend to prioritize quantity features over time-based ones. This may reflect an increased sensitivity to DoS attacks, in which attackers generate a high volume of flows over short periods. The detailed SHAP values are presented in Table~\ref{tab:multi-shap}.

\begin{table}[h]
\caption{Multiclass feature importance}
\label{tab:multi-shap}
\begin{center}
\begin{tabular}{@{}llll@{}}
\toprule
\textbf{Model} & \textbf{Quantity-based} & \textbf{Time-based} & \textbf{Hybrid-based} \\ \midrule
RF   & 2.87 & 0.86 & 0.28 \\
XGB  & 3.50  & 0.22 & 0.27 \\
LGBM & 3.35 & 0.38 & 0.27 \\
LSTM & 2.64 & 1.19 & 0.16 \\
MLP  & 2.84 & 0.90 & 0.26 \\ \bottomrule
\end{tabular}
\end{center}
\end{table}

\subsection{Analysis and discussion}
\label{analysis}

Overall, all binary and multiclass models achieved high performance metrics. While a quarter of the features in the smaller subset differed between the two classification objectives, the models maintained most of their behavioral patterns. Using this reduced feature set slightly affected prediction performance, while reducing TT and inference time IT.

While binary models attained higher F1S, multiclass models consistently maintained a lower FPR for both the full and smaller feature sets. This suggests that, while multiclass models are better at distinguishing benign flows from attacks, they have more difficulty distinguishing between specific types of attacks. Nonetheless, given that the selected feature subsets focused on time, quantity, and hybrid metrics while excluding topology-related features, the models' near-perfect results, particularly in binary classification, demonstrate their ability to successfully generalize from statistical flow behaviors and accurately classify unseen data.

When using the full feature set for binary classification, RF and XGB had similar performance, even though XGB required less TT and IT. With the smaller set, RF and LGBM had an equal F1S, but LGBM required only one-tenth the TT and half the IT, matching XGB’s IT. Among the neural networks, the MLP achieved a lower FPR than the LSTM while requiring less TE. However, with the reduced feature set, LSTM achieved a higher F1S and lower FPR, despite having the longest TE and IT compared to MLP.

Overall, in multiclass classification, the simpler RF model was the top-performing model, achieving the highest F1S and lowest FPR with both the full and smaller feature sets. Even when using fewer features, RF had the highest F1S, although its FPR matched that of LGBM with both feature sets. LGBM required the least amount of TT, and XGB offered the fastest inference time. Though the DL models did not reach the ML models’ top scores, they achieved comparable results and exhibited the highest FPR, especially with the smaller subset.

\section{Conclusions}
\label{conclusion}

This study presented an experimental validation and evaluation of the reliability of the GeNIS dataset for AI-based NIDS, to serve as a baseline for future benchmarks. The dataset was carefully analyzed, and multiple feature selection techniques were combined to identify the most relevant features and prepare them for the training of ML and DL models. 

A subset of 16 features, accounting for approximately 70\% of the total feature importance, was selected by combining the outputs of these methods. These features were decisive in distinguishing between benign and malicious flows, as well as between different types of malicious activity. In the binary classification task, the selected features were roughly balanced between quantity, time, and hybrid-based types. In the multiclass task, however, quantity features were four times more frequent and time-based features were three times more frequent than hybrid ones.

The RF, XGB, LGBM, LSTM, and MLP models obtained high results across both binary and multiclass classification tasks. The ensemble models consistently achieved higher scores and required less training and inference time with both the full feature set and the reduced subset. Since the reduced feature set only slightly impacted performance and was composed of behavior-based features rather than topology-specific ones, the results suggest good generalization potential to other datasets. SHAP analysis revealed that the models predominantly relied on quantity-based features, which may indicate an overreliance on traffic volume or packet counts as key indicators of malicious behavior.

It is important to continue the research and development efforts to create larger and more diverse network traffic datasets and make them publicly available. In the future, as novel datasets are created, a promising approach will be the combination of multiple datasets and the investigation of the transferability of the detection capabilities of ML and DL models to previously unseen network traffic. Furthermore, given the high generalization results of all models, it is crucial to extend this study to assess model robustness in adversarial settings as a step towards real-world deployment.


\section*{Acknowledgment}

 This work was supported by the PC2phish project, which has received funding from FCT with Refª: 2024.07648.IACDC. This work has also received funding from UIDB/00760/2020.


\bibliographystyle{splncs04}
\bibliography{bibliography}

@article{traditional_ids,
  title = {A Review of Enhancing Intrusion Detection Systems for Cybersecurity Using Artificial Intelligence (AI)},
  volume = {29},
  ISSN = {2451-3113},
  DOI = {10.2478/kbo-2023-0072},
  number = {3},
  journal = {International conference KNOWLEDGE-BASED ORGANIZATION},
  publisher = {Walter de Gruyter GmbH},
  author = {Markevych,  Michal and Dawson,  Maurice},
  year = {2023},
  month = jun,
  pages = {30–37}
}

@ARTICLE{benchmark_cicids2017,
  author={Maseer, Ziadoon and others},
  journal={IEEE Access}, 
  title={Benchmarking of Machine Learning for Anomaly Based Intrusion Detection Systems in the CICIDS2017 Dataset}, 
  year={2021},
  volume={9},
  number={},
  pages={22351-22370},
  doi={10.1109/ACCESS.2021.3056614}}

@inproceedings{benchmark_cicids2017_2,
author = {Catillo, Marta and others},
title = {A Case Study with CICIDS2017 on the Robustness of Machine Learning against Adversarial Attacks in Intrusion Detection},
year = {2023},
isbn = {9798400707728},
publisher = {Association for Computing Machinery},
address = {New York, NY, USA},
doi = {10.1145/3600160.3605031},
booktitle = {Proceedings of the 18th International Conference on Availability, Reliability and Security},
articleno = {74},
numpages = {8},
location = {Benevento, Italy},
series = {ARES '23}
}

@article{PINTO2025111177,
title = {A review on intrusion detection datasets: tools, processes, and features},
journal = {Computer Networks},
volume = {262},
pages = {111177},
year = {2025},
issn = {1389-1286},
doi = {10.1016/j.comnet.2025.111177},
author = {Daniela Pinto and others}
}

@conference{cicids,
author={Iman Sharafaldin and others},
title={Toward Generating a New Intrusion Detection Dataset and Intrusion Traffic Characterization},
booktitle={Proceedings of the 4th International Conference on Information Systems Security and Privacy - ICISSP},
year={2018},
pages={108-116},
publisher={SciTePress},
organization={INSTICC},
doi={10.5220/0006639801080116},
isbn={978-989-758-282-0},
issn={2184-4356},
}

@INPROCEEDINGS{engelen,
  author={Engelen, Gints and others},
  booktitle={2021 IEEE Security and Privacy Workshops (SPW)}, 
  title={Troubleshooting an Intrusion Detection Dataset: the CICIDS2017 Case Study}, 
  year={2021},
  volume={},
  number={},
  pages={7-12},
  doi={10.1109/SPW53761.2021.00009}}

@INPROCEEDINGS{engelen_2,
  author={Liu, Lisa and others},
  booktitle={2022 IEEE Conference on Communications and Network Security (CNS)}, 
  title={Error Prevalence in NIDS datasets: A Case Study on CIC-IDS-2017 and CSE-CIC-IDS-2018}, 
  year={2022},
  volume={},
  number={},
  pages={254-262},
  doi={10.1109/CNS56114.2022.9947235}}

@InProceedings{Lanvin,
author="Lanvin, Maxime
and others",
title="Errors in the CICIDS2017 Dataset and the Significant Differences in Detection Performances It Makes",
booktitle="Risks and Security of Internet and Systems",
year="2023",
publisher="Springer Nature Switzerland",
address="Cham",
pages="18--33",
isbn="978-3-031-31108-6"
}

@inproceedings{rosay,
author = {Rosay, Arnaud and others},
title = {From CIC-IDS2017 to LYCOS-IDS2017: A corrected dataset for better performance},
year = {2022},
isbn = {9781450391153},
publisher = {Association for Computing Machinery},
address = {New York, NY, USA},
doi = {10.1145/3486622.3493973},
booktitle = {IEEE/WIC/ACM International Conference on Web Intelligence and Intelligent Agent Technology},
pages = {570–575},
numpages = {6},
location = {Melbourne, VIC, Australia},
series = {WI-IAT '21}
}

@Article{hikari,
AUTHOR = {Ferriyan, Andrey and others},
TITLE = {Generating Network Intrusion Detection Dataset Based on Real and Encrypted Synthetic Attack Traffic},
JOURNAL = {Applied Sciences},
VOLUME = {11},
YEAR = {2021},
NUMBER = {17},
ARTICLE-NUMBER = {7868},
ISSN = {2076-3417},
DOI = {10.3390/app11177868}
}

@INPROCEEDINGS{fernandes,
  author={Fernandes, Rui and others},
  booktitle={2023 11th International Symposium on Digital Forensics and Security (ISDFS)}, 
  title={The impact of identifiable features in ML Classification algorithms with the HIKARI-2021 Dataset}, 
  year={2023},
  volume={},
  number={},
  pages={1-5},
  doi={10.1109/ISDFS58141.2023.10131864}}

@dataset{new_hikari,
  author       = {Ferriyan, Andrey and
                  others},
  title        = {HIKARI-2021: Generating Network Intrusion
                   Detection Dataset Based on Real and Encrypted
                   Synthetic Attack Traffic
                  },
  month        = apr,
  year         = 2022,
  publisher    = {Zenodo},
  version      = {1.4.0},
  doi          = {10.5281/zenodo.6463389},
}

@inproceedings{Hera2025,
  author={Pinto, Daniela and others},
  booktitle={2024 IEEE 23rd International Conference on Trust, Security and Privacy in Computing and Communications (TrustCom)}, 
  title={A Novel Approach to Network Traffic Analysis: the HERA tool}, 
  year={2024},
  volume={},
  number={},
  pages={1850-1856},
  doi={10.1109/TrustCom63139.2024.00255}}

@article{genis,
title = {GeNIS: A modular dataset for network intrusion detection and classification},
journal = {Data in Brief},
volume = {60},
pages = {111487},
year = {2025},
issn = {2352-3409},
doi = {10.1016/j.dib.2025.111487},
author = {Miguel Silva and others}
}

@article{lstm_rw,
	author = {Jinsi Jose and Deepa Jose},
	title = {Deep learning algorithms for intrusion detection systems in internet of things using CIC-IDS 2017 dataset},
	journal = {International Journal of Electrical and Computer Engineering (IJECE)},
	volume = {13},
	number = {1},
	year = {2023},
	issn = {2722-2578},	
        pages = {1134--1141},	
        doi = {10.11591/ijece.v13i1.pp1134-1141}
}

@article{lstm_rw2,
  title = {Intrusion detection systems using long short-term memory (LSTM)},
  volume = {8},
  ISSN = {2196-1115},
  DOI = {10.1186/s40537-021-00448-4},
  number = {1},
  journal = {Journal of Big Data},
  publisher = {Springer Science and Business Media LLC},
  author = {Laghrissi,  FatimaEzzahra and others},
  year = {2021},
  month = may 
}

@article{arvores,
  title = {Reliable feature selection for adversarially robust cyber-attack detection},
  volume = {80},
  ISSN = {1958-9395},
  DOI = {10.1007/s12243-024-01047-z},
  number = {3–4},
  journal = {Annals of Telecommunications},
  publisher = {Springer Science and Business Media LLC},
  author = {Vitorino,  João and others},
  year = {2024},
  month = jun,
  pages = {341–355}
}

@InProceedings{arvores2,
author="Vitorino, Jo{\~a}o
and others",
title="An Adversarial Robustness Benchmark for Enterprise Network Intrusion Detection",
booktitle="Foundations and Practice of Security",
year="2024",
publisher="Springer Nature Switzerland",
address="Cham",
pages="3--17",
isbn="978-3-031-57537-2"
}

@article{mlp,
  title = {IGRF-RFE: a hybrid feature selection method for MLP-based network intrusion detection on UNSW-NB15 dataset},
  volume = {10},
  ISSN = {2196-1115},
  DOI = {10.1186/s40537-023-00694-8},
  number = {1},
  journal = {Journal of Big Data},
  publisher = {Springer Science and Business Media LLC},
  author = {Yin,  Yuhua and others},
  year = {2023},
  month = feb 
}

@article{information_gain,
  title = {Determining threshold value on information gain feature selection to increase speed and prediction accuracy of random forest},
  volume = {8},
  ISSN = {2196-1115},
  DOI = {10.1186/s40537-021-00472-4},
  number = {1},
  journal = {Journal of Big Data},
  publisher = {Springer Science and Business Media LLC},
  author = {Prasetiyowati,  Maria Irmina and others},
  year = {2021},
  month = jun 
}

@inproceedings{chi_squared,
  title = {Chi-Squared Based Feature Selection for Stroke Prediction using AzureML},
  DOI = {10.1109/ietc47856.2020.9249117},
  booktitle = {2020 Intermountain Engineering,  Technology and Computing (IETC)},
  publisher = {IEEE},
  author = {Ray,  Sujan and others},
  year = {2020},
  month = oct,
  pages = {1–6}
}

@INPROCEEDINGS{rfe,
  author={Sachdeva, Ravi and others},
  booktitle={2022 2nd International Conference on Advance Computing and Innovative Technologies in Engineering (ICACITE)}, 
  title={A Systematic Method for Breast Cancer Classification using RFE Feature Selection}, 
  year={2022},
  volume={},
  number={},
  pages={1673-1676},
  doi={10.1109/ICACITE53722.2022.9823464}}

@article{mad,
title = {Feature selection based on absolute deviation factor for text classification},
journal = {Information Processing \& Management},
volume = {60},
number = {3},
pages = {103251},
year = {2023},
issn = {0306-4573},
doi = {10.1016/j.ipm.2022.103251},
author = {Lingbin Jin and others}
}

@article{dr,
title = {Dispersion Ratio based Decision Tree Model for Classification},
journal = {Expert Systems with Applications},
volume = {116},
pages = {1-9},
year = {2019},
issn = {0957-4174},
doi = {10.1016/j.eswa.2018.08.039},
author = {Smita Roy and others}
}

@article{lstm_author,
    author = {Hochreiter, Sepp and Schmidhuber, Jürgen},
    title = {Long Short-Term Memory},
    journal = {Neural Computation},
    volume = {9},
    number = {8},
    pages = {1735-1780},
    year = {1997},
    month = {11},
    issn = {0899-7667},
    doi = {10.1162/neco.1997.9.8.1735},
}

@inproceedings{lgbm_author,
 author = {Ke, Guolin and others},
 booktitle = {Advances in Neural Information Processing Systems},
 pages = {},
 publisher = {Curran Associates, Inc.},
 title = {LightGBM: A Highly Efficient Gradient Boosting Decision Tree},
 volume = {30},
 year = {2017}
}

@article{rf_author,
author={Breiman, Leo},
title={Random Forests},
journal={Machine Learning},
year={2001},
month={Oct},
day={01},
volume={45},
number={1},
pages={5-32},
issn={1573-0565},
doi={10.1023/A:1010933404324},
}

@inproceedings{xgb_author,
author = {Chen, Tianqi and Guestrin, Carlos},
title = {XGBoost: A Scalable Tree Boosting System},
year = {2016},
isbn = {9781450342322},
publisher = {Association for Computing Machinery},
address = {New York, NY, USA},
doi = {10.1145/2939672.2939785},
booktitle = {Proceedings of the 22nd ACM SIGKDD International Conference on Knowledge Discovery and Data Mining},
pages = {785–794},
numpages = {10},
location = {San Francisco, California, USA},
series = {KDD '16}
}

@article{mlp_author,
  title = {The perceptron: A probabilistic model for information storage and organization in the brain.},
  volume = {65},
  ISSN = {0033-295X},
  DOI = {10.1037/h0042519},
  number = {6},
  journal = {Psychological Review},
  publisher = {American Psychological Association (APA)},
  author = {Rosenblatt,  F.},
  year = {1958},
  pages = {386–408}
}

@inproceedings{shap,
author = {Lundberg, Scott and Lee, Su-In},
title = {A unified approach to interpreting model predictions},
year = {2017},
isbn = {9781510860964},
publisher = {Curran Associates Inc.},
address = {Red Hook, NY, USA},
booktitle = {Proceedings of the 31st International Conference on Neural Information Processing Systems},
pages = {4768–4777},
numpages = {10},
location = {Long Beach, California, USA},
series = {NIPS'17}
}

@INPROCEEDINGS{outdated,
  author={Hnamte, Vanlalruata and Hussain, Jamal},
  booktitle={2021 3rd International Conference on Electrical, Control and Instrumentation Engineering (ICECIE)}, 
  title={An Extensive Survey on Intrusion Detection Systems: Datasets and Challenges for Modern Scenario}, 
  year={2021},
  volume={},
  number={},
  pages={1-10},
  doi={10.1109/ICECIE52348.2021.9664737}}

@ARTICLE{flaws,
  author={Khanan, Akbar and others},
  journal={IEEE Access}, 
  title={From Bytes to Insights: A Systematic Literature Review on Unraveling IDS Datasets for Enhanced Cybersecurity Understanding}, 
  year={2024},
  volume={12},
  number={},
  pages={59289-59317},
  doi={10.1109/ACCESS.2024.3392338}}

@ARTICLE{outdated2,
  author={Janabi, Ahmed and others},
  journal={IEEE Access}, 
  title={Survey: Intrusion Detection System in Software-Defined Networking}, 
  year={2024},
  volume={12},
  number={},
  pages={164097-164120},
  doi={10.1109/ACCESS.2024.3493384}}

@INPROCEEDINGS{hikari_22_analysis,
  author={Rahman, Md. and others},
  booktitle={2024 27th International Conference on Computer and Information Technology (ICCIT)}, 
  title={Enhancing Network Intrusion Detection with Deep Learning: A Comprehensive Analysis}, 
  year={2024},
  volume={},
  number={},
  pages={1028-1033},
  doi={10.1109/ICCIT64611.2024.11022353}}

\end{document}